# An open-source platform to study uniaxial stress effects on nanoscale devices

G. Signorello, M. Schraff, P. Zellekens[$], U. Drechsler, M. Buerge, H.R. Steinauer, R. Heller, M. Tschudy, and H. Riel

*IBM Research – Zurich, 8803 Rüschlikon, Switzerland*

[$] *Current address:* Peter Grünberg Institute, Forschungszentrum Jülich GmbH, 52425 Jülich, Germany.

Corresponding author email: giorgio.signorello@gmail.com

## Abstract

We present an automatic measurement platform that enables the characterization of nanodevices by electrical transport and optical spectroscopy as a function of uniaxial stress. We provide insights into and detailed descriptions of the mechanical device, the substrate design and fabrication, and the instrument control software, which is provided under open-source license. The capability of the platform is demonstrated by characterizing the piezo-resistance of an InAs nanowire device using a combination of electrical transport and Raman spectroscopy. The advantages of this measurement platform are highlighted by comparison with state-of-the-art piezo-resistance measurements in InAs nanowires. We envision that the systematic application of this methodology will provide new insights into the physics of nanoscale devices and novel materials for electronics, and thus contribute to the assessment of the potential of strain as a technology booster for nanoscale electronics.

## Introduction

The effect of mechanical stress on the electronic properties semiconductors is a well-established field of physics[1,2] and has seen renewed interest in view of its applications to nanoscale materials. Strain allows the manipulation of the band structure to render specific bands energetically favorable and to modify the distribution of charge carriers and the effective mass of the bands. This effect has been used to enhance the performance of complementary metal oxide semiconductor (CMOS) field-effect transistors and to advance optoelectronic devices based on III-V alloys.[3–10]

Semiconducting nanostructures also exhibit interesting mechanical properties when their lateral dimension are scaled below 100 nm.[11] Because of the competition between atomic coordination and electronic distribution, surfaces can be softer or harder than the bulk.[12] As a consequence, mechanical properties like Young's modulus can be increased[13] or decreased[14] compared with those of bulk crystals by decreasing the nanowire cross section down to the nanoscale. By leveraging this class of phenomena, high values of strain were obtained in semiconducting nanostructures.[15] The enlarged range of elastic strain of the nanostructures revealed remarkable piezo-resistance properties in group-IV semiconductors and III-V alloys,[16–19] and was used to explore new limits for tunable light-emitting nanostructures,[20] observe novel band-structure transitions,[21,22] create novel device concepts for energy harvesting,[23] and investigate reliability aspects of nanomaterials for energy storage.[24,25]

Various instruments have enabled these scientific explorations and discoveries. The diamond anvil cell (DAC) has been the workhorse of high-pressure physics. By allowing to achieve large values of hydrostatic or shear stress, this instrument enabled the study of their influence on the optical properties, lattice dynamics and electronic transport of bulk samples and thin films.[26,27] Using the DAC, the optical properties and lattice dynamics of nanostructures have been studied,[28,29] but insights into how stress influences the electronic transport properties of these nanostructures remains limited. Bending mechanics, using three-point or four-point geometries, has been the concept of choice to study the effect of strain on the electronic transport of thin films, grown epitaxially on III-V substrates or silicon.[30–32] Electrical properties are easily accessed, and the applied stress can be determined using strain gauges or continuum-mechanics considerations on the bending geometry of the substrate. Silicon and III-V alloys, however, cannot accommodate large stress upon bending: cracks are created on the substrate surface and propagate, causing brittle failure of the substrate. For this reason, the stress applicable to traditional semiconducting substrates does not exceed few hundreds of GPa, and the corresponding strain values are smaller than 1%.

To determine the local stress in nanostructures, tensile test machines were realized by means of bulk silicon micromachining and used in conjunction with scanning electron microscopy, transmission electron microscopy or X-ray diffraction. These experiments revealed remarkable mechanical properties and modifications of the electronic properties of nanotubes[33,34] as well as of metallic[35,36] and semiconducting nanowires.[19,37,38] Those methods, however, suffer from a limited freedom of choice of the materials used to establish electrical contact with the nanostructure. As a consequence, the effect of strain on the electronic properties of the device is often dominated by Schottky contacts and by the influence that strain has on the barrier height through the piezo-electric effect.[39]

In this work, we describe a measurement platform that enables the characterization of nanodevices by means of electrical transport and optical spectroscopy as a function of uniaxial stress. The platform relies on standard semiconductor fabrication methods to establish contact to nanostructures, and ultimately is able to reveal the effect large values of stress have on the transport properties of the device. Nanowire devices are fabricated on flexible and compliant substrates. Bending the substrate in a three-point geometry induces an extension or compression of the substrate surface, and a part of this strain is transferred the nanostructure under test by metal clamps. Raman spectroscopy is used to determine the strain tensor in the structure. The electrical transport properties are characterized while stress is being applied, without interfering with the optics. We provide insights into and detailed descriptions of the bending mechanics, the substrate design and fabrication, and the instrument control software, which is provided under open-source license. We demonstrate the capability of the platform by characterizing the piezo-resistance of an InAs nanowire. After describing the sample fabrication, we show how the two-probe and four-probe characteristics of the nanowire vary under the application of stress and correlate those findings with the changes in the Raman spectrum.

# Strain and Characterization Setup

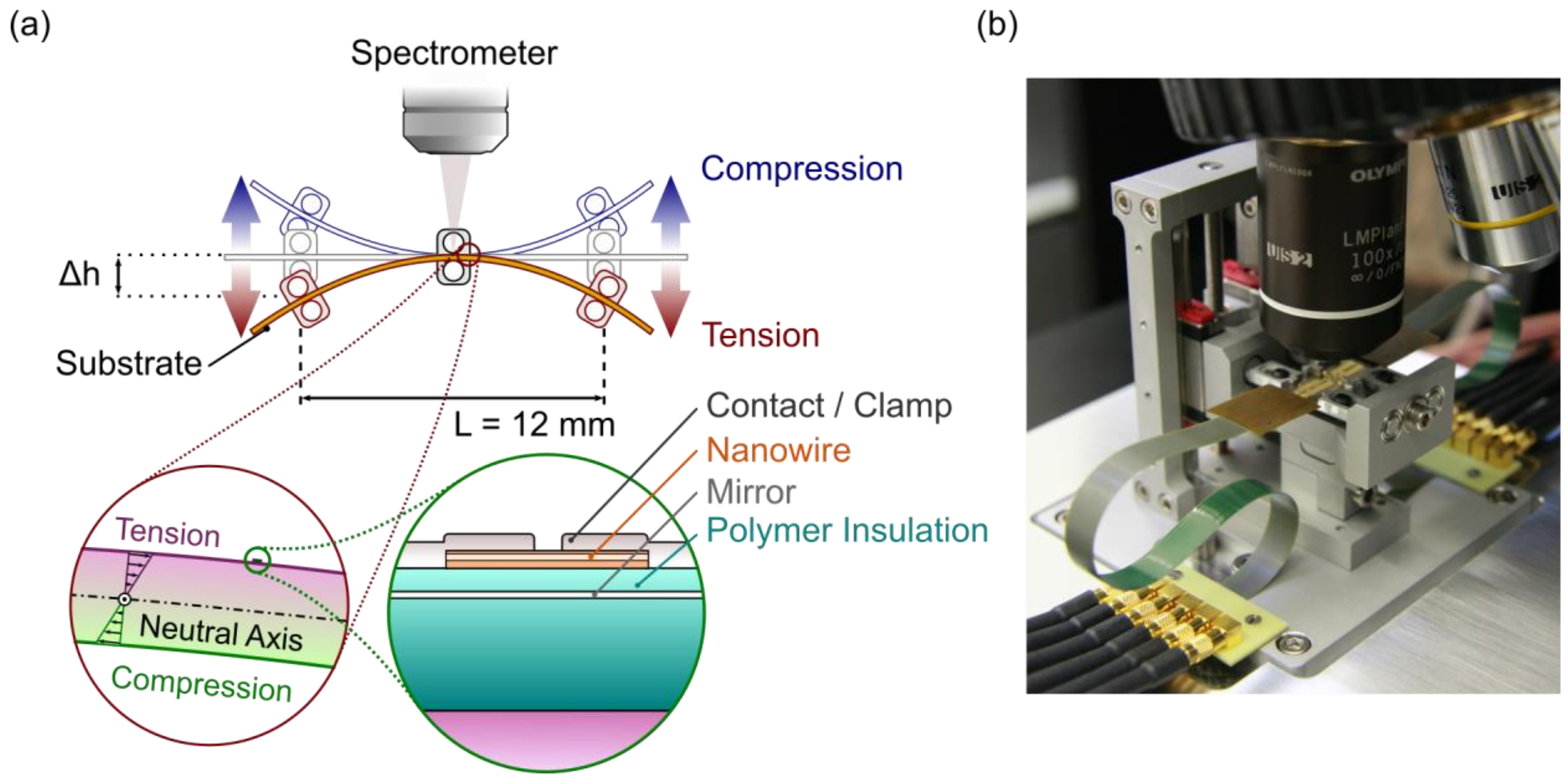

Figure 1: (a) Schematic representation of the experimental setup used to perform the electrical and optical characterization of nanoscale devices as a function of strain. (b) Photo of the bending mechanism realized for the platform. Electrical contact is established on the sides of the bending chip using off-the-shelf connectors, which move together with the chip.

## Bending Mechanics

The setup and the principles that govern its operation are sketched in Figure 1(a). The substrate is attached to the mechanism in three points: one central clamp and two lateral supports, which are mounted to a common base with roller bearings and are free to rotate. The distance between the central clamp and the lateral support base is controlled with a differential thread, i.e., a thread consisting of two segments of different diameter and pitch. As the thread rotates under the action of a rotation stage (Newport SR50), the base to which the lateral supports are connected moves in the vertical direction relative to the block on which the central clamp is attached. Depending on the direction of travel, the substrate is bent in a convex or a concave fashion, and its surface experiences tensile or compressive strain, respectively. Using continuum-mechanics considerations,[40–42] one can show that the strain in the substrate surface is proportional to the local bending moment, which is maximum at the center and decreases linearly to zero at the lateral supports. Thus, the device under test must be located close to the central clamp to couple the highest possible strain to it. The substrate-bending mechanics allows a strain range of up to 10%, with an accuracy on the order of one part in $10^9$, without inducing plastic deformation in the substrate. The nanowire experiences a bending smaller than $8\times10^{-5}$ °/%: the mechanism induces a purely uniaxial stress on the nanowire.[42]

## Optical Spectrometer

A high-numerical-aperture objective with a long working distance of 3.5 mm (Olympus LMPlanFL 100X) permits a good optical collection efficiency and a large travel range without collision between the objective and the electrical connectors. The objective is close to the center of the beam to capture the optical spectra of the device under test. The central clamp is maintained at a fixed position to limit the changes in working distance induced by bending and to facilitate the focusing procedure.

The Raman spectra are collected by a commercial optical spectrometer, a Labram HR by Horiba Scientific. The laser excitation, emitted by a $Nd:YVO_4$ diode-pumped laser at 532 nm, can be attenuated using a neutral density filter and is reflected to the objective using a long pass filter. The scattered light from the device under test is collected by the objective, transmitted by the long pass filter, focused onto a confocal hole, and diffracted by a grating with 1800 lines per millimeter and blazing angle optimized for 850 nm. Finally, it is detected by a liquid-nitrogen-cooled CCD (silicon Horiba Symphony open electrode CCD).

## Electrical Measurements and Contacts

For electrical measurements of the devices, multiple electrical contacts must to be established on the substrate surface and maintained upon bending, while avoiding collisions with the optics. Our solution establishes the contacts on the edges of the chip, beyond the lateral supports. The electrical contact to the substrate is established with a micro sim-card connector mounted on a rigid-flex PCB. With this solution, the connector and the lateral section of the substrate move together upon bending. A photo of the mechanics and of the solution for electrical connection is shown in Figure 1(b). The electrical measurements are performed using a semiconductor parameter analyzer, and the contacts of a single device are connected to the source-measurement units (SMU) using a switching matrix.

# Substrate and Device Fabrication

## Materials Considerations

To enable successful electrical and optical measurements upon bending, the substrate must fulfill several mechanical, optical, and electrical requirements. It must be able to undergo elastic deformations without brittle failure. Stainless steel or beryllium copper disks of 100 mm in diameter and 250 μm in thickness satisfy this property, and therefore have been used as substrate. After cleaning, the substrates are coated with an insulating and compliant polymer: etch-grade Benzo Cyclo-butene (BCB) was chosen because of its compatibility with microfabrication, and its high mechanical compliance, high transparency, low stray luminescence upon laser illumination, and high temperature stability up to 280 °C without any degradation of the mechanical, electrical, and optical properties.

The variation of the contact resistance between substrate and connector was evaluated over multiple stress cycles by measuring the electrical conductivity of metal lines that short-circuit pairs of neighboring contacts. The thickness of the metal used to ensure the contact between the substrate and the connector was critical to the electrical contact reliability: the contact resistance of a 100-nm-thick metal film, which is routinely used to establish contact on silicon substrates, degrades during one stress cycle and leads to loss of contact. Increasing the contact-metal thickness to 2 μm significantly improves the electrical contact reliability. Metal film of this thickness can be realized using an electroplating process. The variation in contact resistance for thick-metal contacts is limited to less than 5 Ω during a cycle that spans the full stress range, from tension to compression and back to the neutral position. This resistance variation is acceptable for the characterization of high-impedance devices.

## Substrate Fabrication

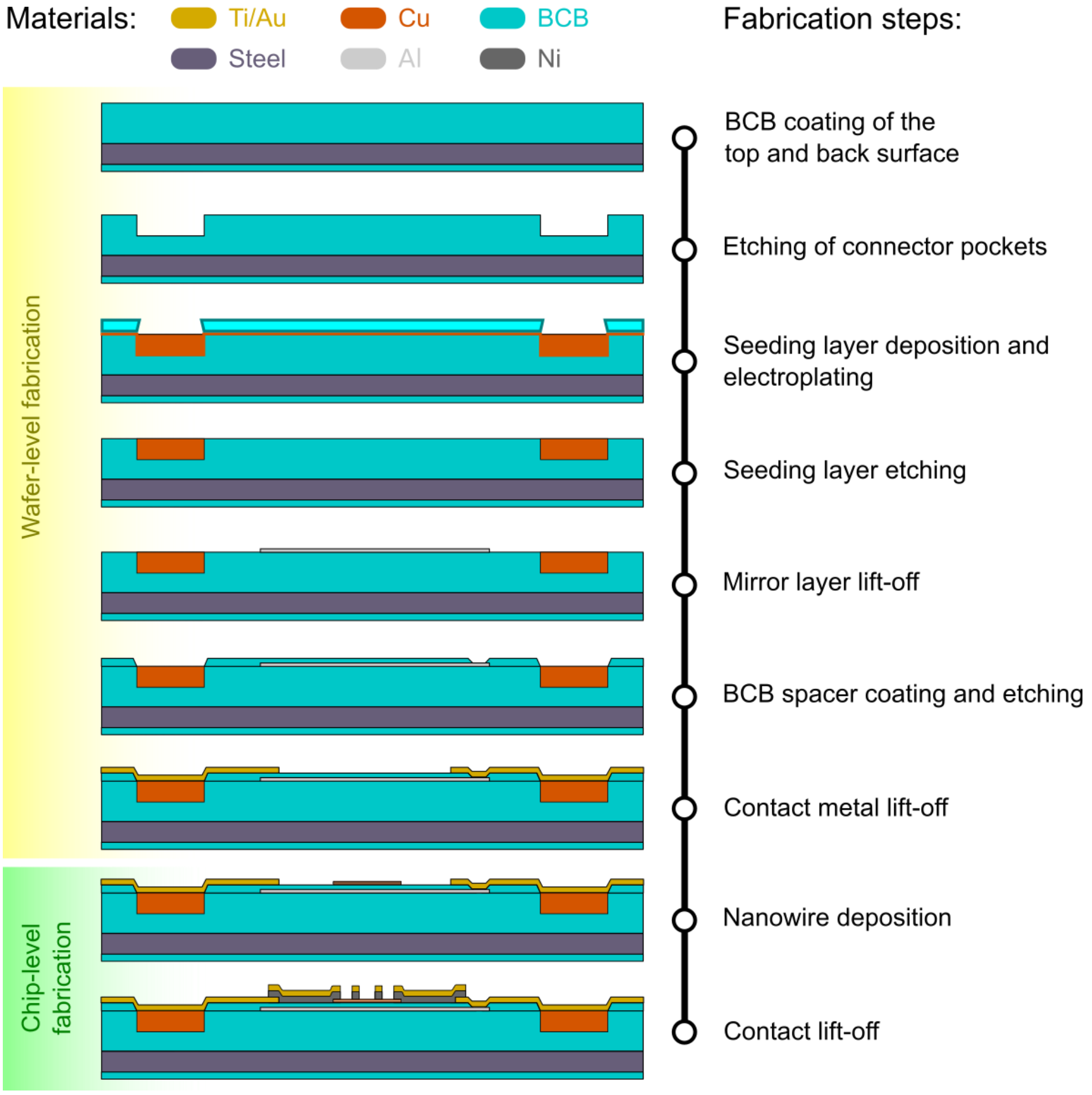

Figure 2: Fabrication procedure for the strain substrate and a simple nanowire device. The flexible substrate is realized using the steps highlighted in yellow. The steps to establish the electrical contact to the nanostructure and to finalize the device are highlighted in green and are performed on individual chips.

The process used to fabricate the nanowire devices is illustrated in Figure 2. The metal substrates are encapsulated by coating the front-side and the back-side with BCB to make its surface electrically insulating. After each spin-coating step, the polymer is cured in a vacuum oven at 210 °C and flushed with nitrogen at a pressure of 400 mbar.[43] The front-side BCB layer has a thickness of about 12 µm, its purpose is to decrease the roughness of the surface to below 5 nm. Pockets are etched into the BCB layer using reactive ion etching (RIE) and a sacrificial mask made by photolithography.[44] The etching step is performed using a mixture of oxygen and sulfur hexafluoride ($SF_6$) at a pressure of 150 mbar and with fluxes of 5 sccm and 50 sccm, respectively. The etching step occurs at 80 W of power and a voltage of 200 V. After that, a 50-nm-thin layer of copper is deposited by sputtering and used as seeding layer for the electroplating. The substrate is then covered with photoresist, and photolithography is used to reveal the pockets defined in the preceding etching step. The pockets are filled selectively with 2 µm of copper using electroplating. After deposition, a short etching step is performed to remove the portion of the seeding layer that is covered by the resist mask and to electrically isolate the copper contacts from one another.[45] Next, a small square of aluminum is realized at the center of each chip using photolithography and liftoff. The introduction of this metal film has two key advantages: it acts as mirror, increasing the coupling efficiency of the optical signal coming from the device under test, and it can potentially also be electrically contacted and used as gate electrode. On top of it, a thin insulation film of about 200 nm is made with BCB diluted with mesithylene. The film is spin-coated and cured in a vacuum oven. Openings are defined using a photoresist mask and etched with RIE to reveal the electroplated contacts and the aluminum mirror. Finally, photolithography and e-beam deposition of 3 nm of titanium and 80 nm of gold and lift-off in acetone are used to define the metal contacts that connect the contact pads at the edge of the sample with the center of each chip. The metal lines have a meander geometry to limit undesired variations in resistance induced by bending the substrate and applying stress.[46] The substrate is cut into single chips using a metal cutter, and the fabrication continues at the chip level to realize the device to be tested.

## Sample Fabrication

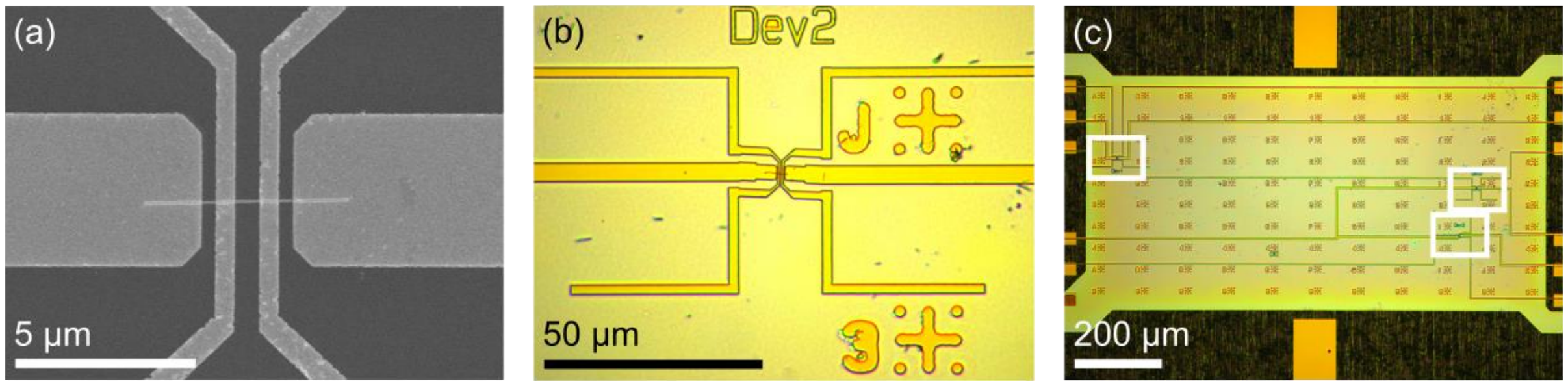

Figure 3: (a) Scanning electron microscope of a single InAs nanowire device; the nanowire is contacted in four-probe configuration. (b) Optical image of an InA nanowire device, contacted in four-probe configuration. The reference markers are used for alignment. (c) Optical image of a strain chip. Three nanowire devices, highlighted with rectangles, are connected in four-probe configuration.

To demonstrate the capability of the setup, we will present the measurements performed on a device made with an InAs nanowire, shown in Figure 3(a). The InAs nanowires were grown by the vapor-liquid-solid (VLS) method in a metalorganic vapor-phase epitaxy (MOVPE) reactor on InAs (111) substrates, using gold nanoparticles as catalyst material. The metalorganic precursors used were trimethylindium and tertiarybutylarsine, with flows of 1.4 μMol/min and 20 μMol/min, respectively. The growth temperature was kept at 430 °C.

Device fabrication begins by harvesting the nanowires from the growth substrate and transferring them to the flexible chip by rubbing these two surfaces in sequence with a piece of cleanroom paper. The sample is then coated with 400 nm of e-beam resist, which is a solution of PMMA 950K diluted in Anisole with 4 % concentration. The sample is baked on a hotplate at 180 °C to allow the solvents to evaporate. The nanowires that are oriented parallel to the length of the chip are imaged with optical microscope, and the images are used to locate the nanowires relative to metal markers, as shown in Figure 3(b). Electron-beam lithography is performed with a beam voltage of 20 kV, a current of 35 pA, and a dose of 270 $\mu C/cm^2$. After exposure, the sample is developed in a solution of Methyl-Isobutyl-Ketone (MIBK) and Isopropanol (IPA) in 1:3 ratio for up to 60 s, and rinsed in IPA. An oxygen-plasma step, performed at 600 W for 15 s with an oxygen flow of 300 sccm, removes possible residues of PMMA and organic contaminants in the contact region of the nanowire. Electron-beam evaporation is used to deposit 80 nm of nickel and 40 nm of gold. The deposition chamber is kept at a pressure of $3\times10^{-7}$ mbar, and the deposition rate is controlled to 0.2 nm/min for both films. The metal that is not in contact with the substrate surface is lifted off in acetone. After that, the sample is rinsed in IPA, and potential traces of acetone absorbed by BCB are evaporated in a convection oven at a temperature 70 °C for 5 min. The contacts have a minimum width of 300 nm and are spaced by gaps of 700 nm in the center and 250 nm on the sides. One strain chip can be used to contact up to three nanowires to the predefined metal contacts, as shown in Figure 3(c).

# Instrument Control Software

## Characterization Procedure

The typical characterization process involves a series of electrical and optical measurements at every step of the applied stress cycle. Electrical transport measurements are performed across each pair of contacts of the device to establish which of the contacts are functional. A small bias voltage of 50 mV is used for the measurement. The device is then measured in four-probe configuration by connecting the outer contacts to two SMUs set as ground and as bias voltage, whereas the inner contacts are connected to two SMUs to measure the voltage drop between them, without draining current. Finally, the nanowire device is measured in two-probe configuration, measuring the current flowing between the two inner contacts.

By defining a sequence of configurations for the parameter analyzer and the switching matrix one can perform the electrical characterization of all devices in a chip. To realize an optical measurement, one must enable the white-light illumination and the spectrometer to capture live images of the device, adjust the working distance of the objective and the position stage so that the device under test is in focus, and compensate for drift and displacements induced by bending the sample. Then, the spectrometer is set in spectroscopy mode by removing the beam splitter used for imaging from the optical path, directing the laser light onto the sample, and performing a spectrum measurement.

Depending on the number of electrical and optical measurements to be performed on each device, the characterization of a chip by electrical transport and optical spectroscopy as function of stress can become a very labor-intensive task. Thus, automating the measurement process can be very beneficial. To do so, we realized a software that enables electrical and optical measurements to be performed automatically as function of stress, and made it available under open-source license.

## Software Architecture

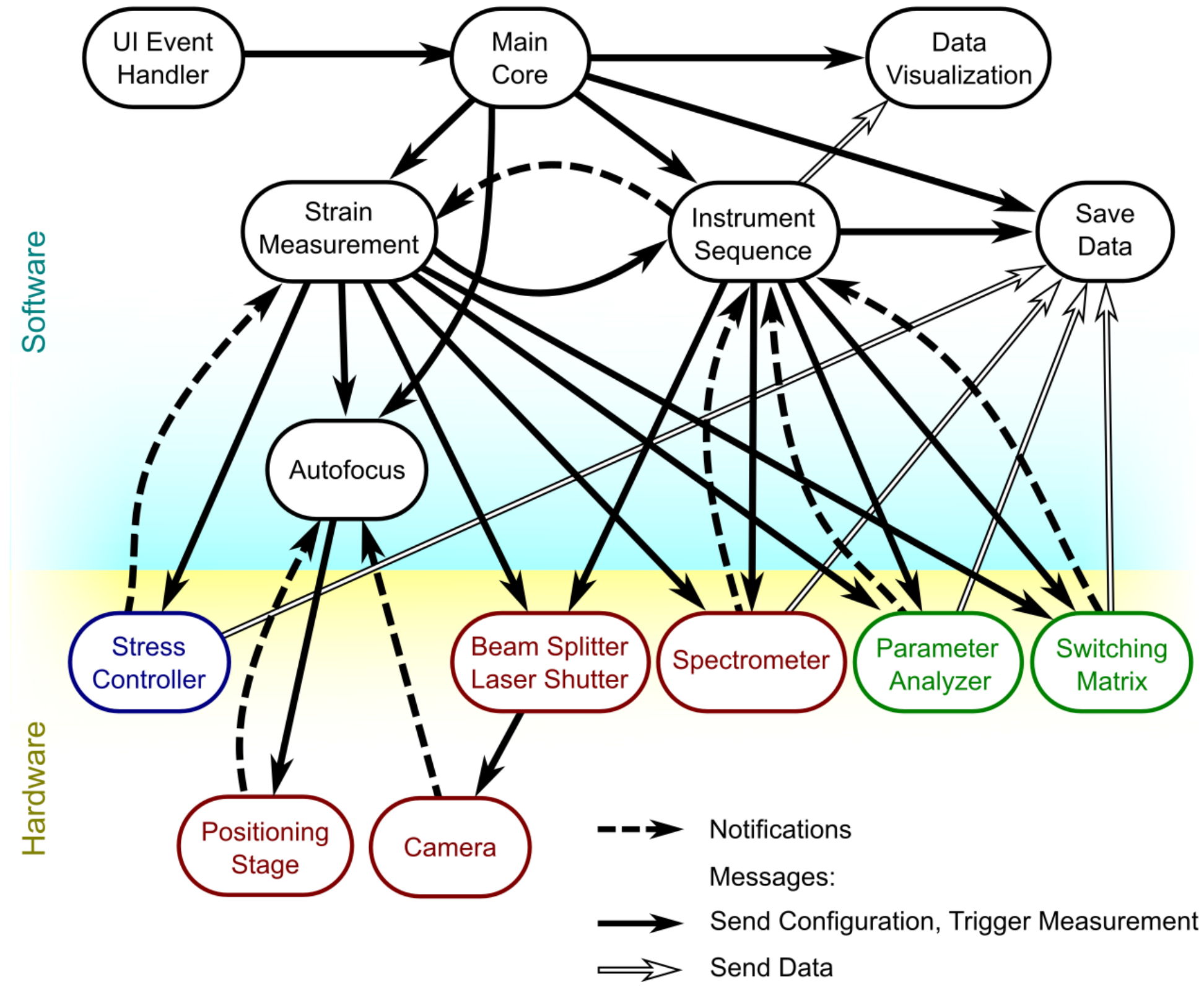

Figure 4: Structure of the instrument control software of the strain setup. Note that to preserve the readability of the diagram, the interactions between the main core (top of the diagram) and the hardware level (bottom of the diagram) have been omitted.

The structure of the software that performs the tasks described above is shown in Figure 4 and has been implemented in Labview using the queued message handler design pattern.[47] The program consists of multiple module cores that run in parallel, represented as bubbles in Figure 4. The module cores execute in response to commands sent as messages (black arrows) to their dedicated queue. When information needs to be accessed by multiple module cores, notifications are used instead (dashed arrows).[48] The execution of a message can trigger a measurement or define a configuration of the instrument, which is stored internally using the functional global variable design pattern.[49] The hardware abstraction level, i.e., the interface to the instruments controlled by the software, is shown in the bottom part of Figure 4. The cores indicated in green represent the electrical measurement apparatus, which consist of the parameter analyzer (Agilent 4155C) and the switching matrix (Keithley 707A). Both instruments are accessed via general-purpose interface bus (GPIB). The rotation stage that controls the bending mechanism, shown in blue, is driven by a controller (Newport ESP301) that is addressed via universal serial bus (USB). The remaining cores, shown in red, represent the optical spectrometer. The camera in the spectrometer (IDS uEye 1540SE) is accessed directly via USB. The positioning stage of the microscope (Märzhäuser Scan75X50) is accessed through serial communication via the controller (Tango Desktop). A relay card (Canakit UK1104), controlled by USB, is used to control the laser shutter and the beam splitter of the Labram HR and to switch between imaging and spectroscopy mode. The interaction with the remaining hardware of the spectrometer is performed via an ActiveX interface to Horiba Labspec 6, which is the control software provided with the spectrometer. Native Labview drivers have been used for all instruments, except for the relay card and the ActiveX interface to Labspec, for which wrappers were made.

The program is executed as follows. First the queues, notifiers and references to the indicators in the user interface are initialized. The initial configurations of the instruments are read from file. The instruments are configured by sending messages to their module cores. Indicators in the front panel of the program signal when the initialization process has been completed and the program is ready to receive commands from the user. All user interactions with the front panel are captured by the user event handler core (top left corner of Figure 4), which sends messages with the details of the user interaction to the main core. In this way, the user interface always remains responsive. Depending on the interaction, the main core can send low-level messages directly to the measurement instruments (bottom of Figure 4, interactions not shown) or it issues high-level commands, such as triggering the autofocus process or starting a sequence of configurations or measurements, to module cores which manage the interaction with the hardware abstraction level. The messages sent to the hardware abstraction level can either configure an instrument to a configuration specified in the message or, as in the case of the spectrometer or the parameter analyzer, trigger a measurement. Once a measurement has been completed, its result is shown to the user directly by dedicated front-panel indicators, which are accessed by the instrument module core by reference. The data measured by each instrument are sent as message (white arrows) to a module core that is responsible for saving the data to file, and all other cores are notified.

The autofocus module implements drift compensation and autofocusing. The user has to define a template image for its operation. This is done by defining a region of interest in a live image of the device, acquired in unstrained conditions before the measurement starts. When in operation, the autofocus module waits for the notification of the latest image captured by the camera. Using a 2D correlation function,[50] the module determines the relative displacement necessary to compensate for the drift in the *x*- and *y*-directions, and sends relative displacement commands to the positioning stage, which notifies it when the respective motion has been completed. To compensate the device focus, the template-matching algorithm is complemented by an edge-detection algorithm to determine the sharpness of the image of the device.[51] This algorithm optimizes the objective working distance to obtain the sharpest live image that resembles the template.

The instrument sequence core allows a list of instrument interactions to be defined and executed. Each interaction can fall into one of two categories:

- Instrument configuration, which consists of configuring the parameter analyzer or correcting the positioning stage with the autofocus routine.
- Launching a measurement, which triggers an electrical measurement, an optical spectrum acquisition, or a combination of both.

The strain measurement core performs the list of configurations and measurements defined in the instrument sequence module for a sequence of strain values that follows a triangular wave-form.

The plot-data core collects the data acquired during an instrument sequence or a strain measurement, and permits the user to browse the data without interfering with the measurement acquisition.

The save data core uses an instance of Python to collect the data and organize it into MATLAB-compatible files for data analysis. The file content differs, depending on whether the measurement is a simple electrical measurement, an optical measurement, a sequence of electrical and optical measurements (henceforth called instrument sequence), or a strain measurement. Electrical measurement data are stored as matrices containing the voltage traces, the current traces, and the time stamp of the measurement for the four SMUs. In contrast, the output of an optical measurement consists of arrays for the spectrum intensity, the wavelength, the energy, and the wavenumber axis. The configurations of the switching matrix, of the parameter analyzer, and of the spectrometer are saved as string. The output of an instrument sequence is a cell-array containing the electrical or optical measurement data at each element and the respective instrument-configuration strings. For a strain measurement, the output consists of an array of elements containing the position of the rotation stage and the corresponding instrument sequence data structure measured at that value of strain.

# Measurement of the effect of uniaxial strain on an InAs nanowire

## Electrical Transport and Raman Characterization

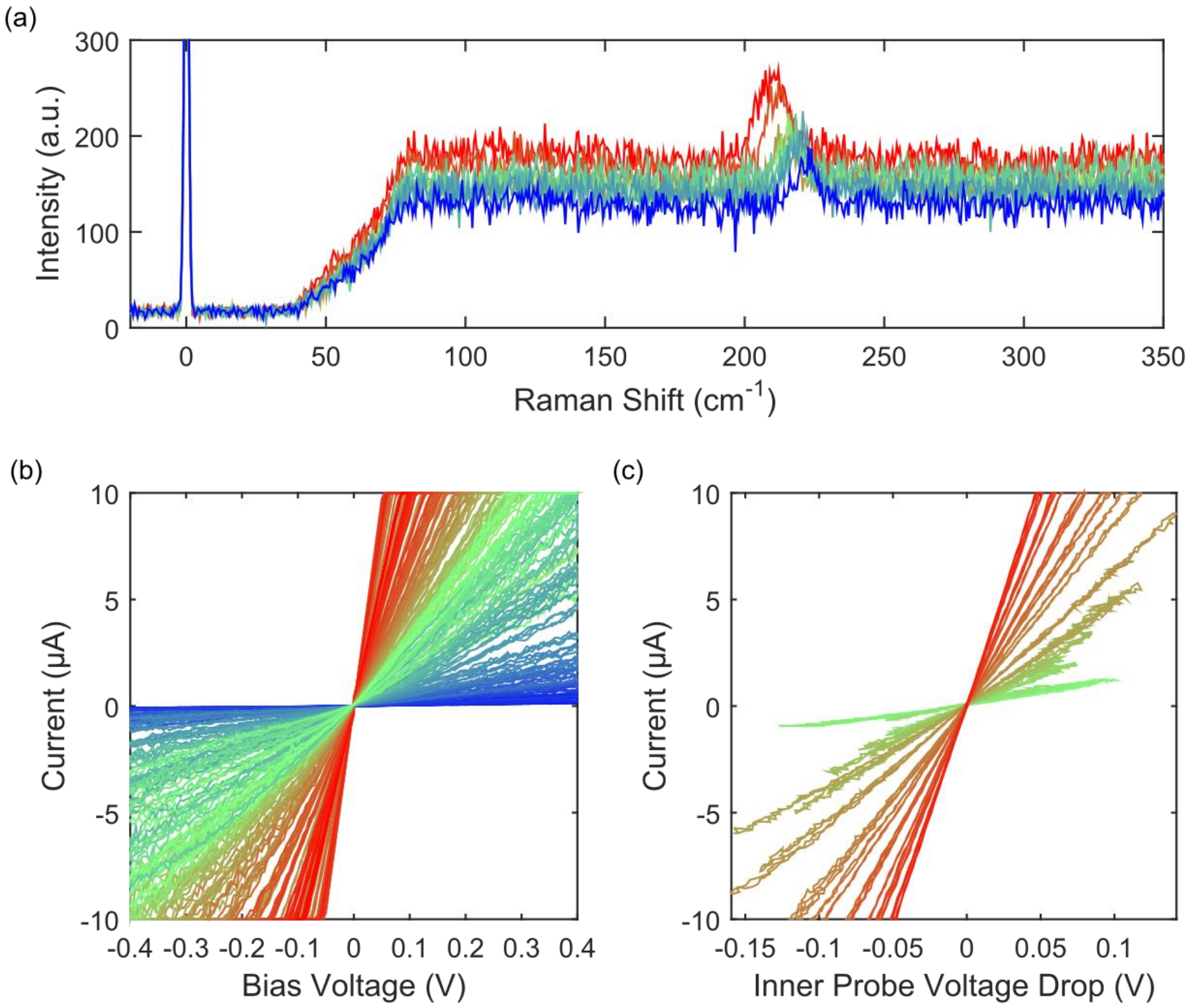

Figure 5: (a) Effect of uniaxial stress on the Raman spectrum of an InAs nanowire device. Data measured in unstrained condition, upon compression and upon tension are shown in green, blue and red, respectively. (b) Effect of uniaxial stress on the two-probe characteristics of an InAs nanowire device, measured between the inner voltage probes for the entire stress cycle. (c) Effect of uniaxial stress on the four-probe characteristics of an InAs nanowire. The current injected between the outer contacts is shown as a function of the voltage drop between the inner leads. This characteristic is not sensitive to contact resistance.

In this section, we present a measurement performed on an InAs nanowire device. The device is subject to a succession of stress values, which follows a triangular wave pattern: starting from the unstrained condition, stress is applied in tension or compression (the direction is set by the user). The stress is increased so that it reaches a maximum value in a sequence of steps, at which electrical and optical characterization is performed. When the maximum stress value has been reached, the motor reverses its motion and heads to maximum stress in the opposite direction. After that, the motor returns the sample back to the unstrained condition.

A selection of the raw electrical and optical data is visualized in Figure 5. A colormap is used to represent the uniaxial stress values that the device experiences. The data points measured in unstrained condition are shown in green, those measured upon compression in blue, and those acquired upon tension in red. The Raman spectra, shown in Figure 5(a), are measured on the central segment of the InAs nanowire using a continuous-wave excitation at 532 nm and 500 µW of power.  Compared with the unstrained condition, the spectra show an increase of the phonon energy upon compression and a decrease upon tension. Strain is known to affect the energy of the optical phonons of semiconductors,[52] and the analysis of the Raman spectrum with stress can be used to determine the strain tensor of the nanowire device.[20,53]

Figure 5(b) shows the two-probe current–voltage (*I*–*V*) characteristics measured at the central segment of the InAs nanowire, and Figure 5(c) summarizes the four-probe characteristics of the nanowire, showing the current injected in the wire between the outer contacts as a function of the voltage drop between the inner contacts. This characteristic is insensitive to the contact resistance of the nanowire, and provides information on the intrinsic electrical transport of the nanowire. Furthermore, the *I*–*V* characteristics of each pair of electrical contacts are measured to identify possible failures of the electrical contacts at each value of applied stress.

In unstrained conditions, the device is fully functional, and the two-probe and four-probe characteristics are measured and shown in green in Figure 5(b, c). When tensile stress is applied, both *I*–*V* characteristics experience an increase in conductivity, shown in red in Figure 5(b, c). Upon maximum tension, one of the current injector contacts fails, and the device therefore can no longer be characterized with the four-probe measurement. The two inner contacts, however, are still intact and maintain their functionality across the entire stress cycle, up to maximum compression. The two-probe electrical characteristics is measured as the device returns to unstrained conditions and upon compression, as shown in blue in Figure 5(b). The electrical resistance increases continuously with increasing compression. This effect is reversed when stress is released from the structure.

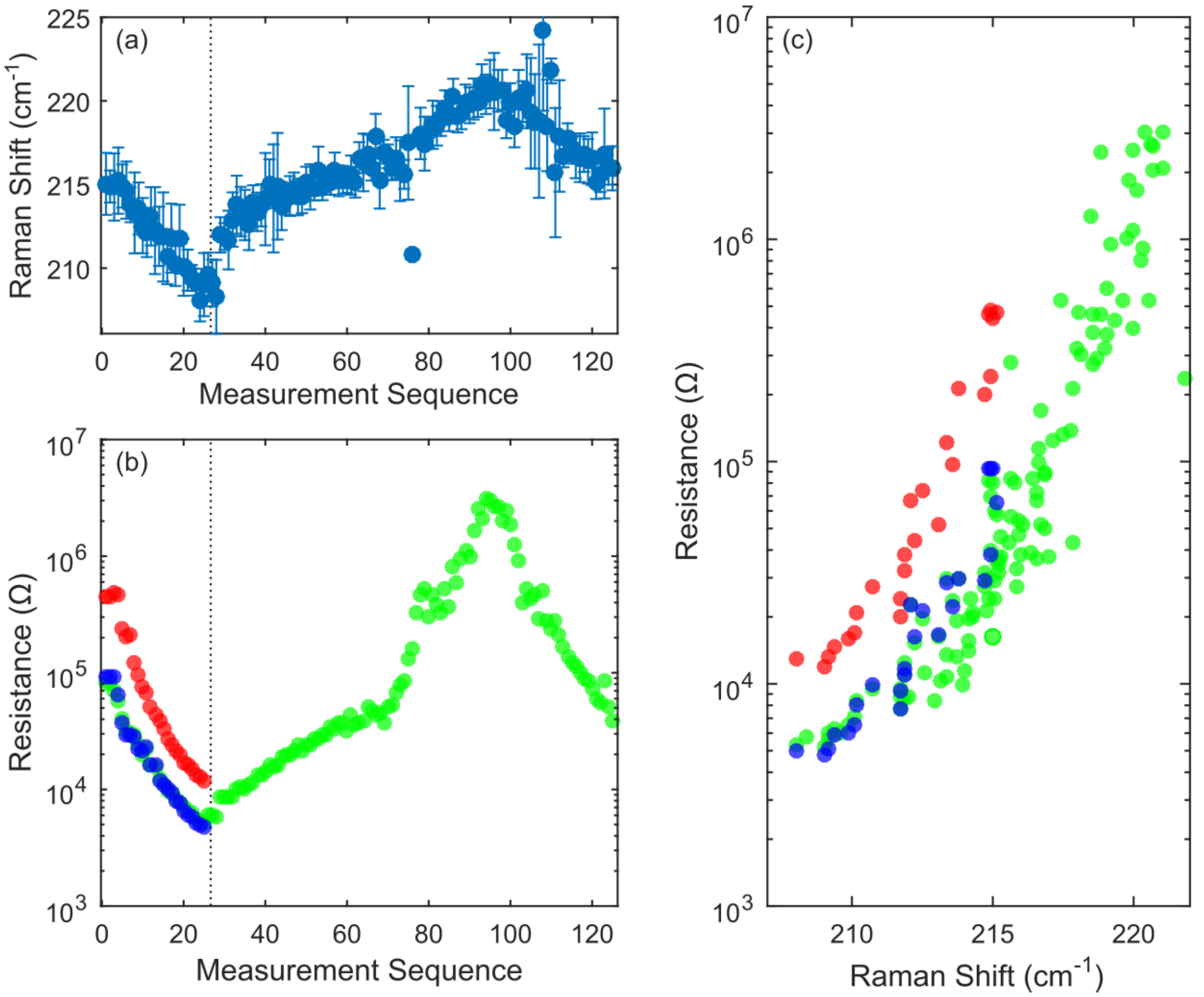

Figure 6: (a) Analysis of the transversal optical (TO) phonon energy shift during a stress cycle. A discontinuity is observed for the tensile-stress value that shifts the TO phonon energy below the threshold of 208 $cm^{-1}$. This threshold corresponds to the failure of the nanowire at one of the outer contacts. (b) Evolution of the four-probe resistance of the InAs device (shown in blue), of the two-probe resistance measured at the inner voltage probes (shown in green), and of the two-probe resistance measured at the outer current injectors (shown in red). The values are shown only in the range for which all contacts necessary to perform a measurement were qualified as functional. (c) The four-probe and two-probe resistances are plotted as function of the InAs TO phonon energy, which can be used to determine the axial strain in the nanowire.

To gain further insights from the measurements, we used a least-squares algorithm to extract the peak position of the Raman spectra, the four-probe resistance, and two-probe resistances for all values of applied stress. The results of the analysis are shown in Figure 6(a, b). At the beginning of the measurement, with increasing tensile stress, the Raman peak shifts linearly to low energy, from 215 $cm^{-1}$ to 208 $cm^{-1}$. All resistances decrease with tensile stress up to a factor of 50. Interestingly, the four-probe resistance and the two-probe resistance measured between the inner contacts have the same value of 100 kΩ in unstrained conditions, and maintain the same value over the entire tensile-stress range. We observe that one of the outer current injection contacts fails when the Raman peak shifts to energies lower than 208 $cm^{-1}$. From this point on, the four-probe resistance can no longer be measured. At the same time, the Raman spectrum undergoes a sudden increase by 5 $cm^{-1}$, which we attribute to the redistribution of mechanical strain within the device after the mechanical failure of one of the contacts.

The two inner contacts maintain their functionality for the remaining steps of the stress measurement sequence, and the two-probe *I–V* characteristics of the central InAs nanowire segment can be measured. The two-probe resistance increases to its unstrained value as stress is reduced to zero and then to higher values with increasing compressive stress. From tension to compression, the two-point resistance is modulated over three orders of magnitude, and the Raman spectrum shifts in energy between 208 $cm^{-1}$ upon tension and 222 $cm^{-1}$ upon compression. When switching from compression to tension, we observe a kink in the slope of the two-probe resistance and the Raman peak position with stress. We believe that such a nonlinear dependence of the electrical transport properties and of the phonon energy with bending is attributed to a mechanical backlash of the bending rig or a warping of the substrate surface. Nevertheless, Figure 6(c) shows that combining the resistance data and the Raman shifts within one plot delivers a curve that is insensitive to the details of the mechanics and that the electron transport properties depend on strain in one continuous relation. This demonstrates the value of Raman as an excellent local strain gauge.

## Discussion

The occurrence of a piezo-resistance observed in InAs upon stress is remarkable and unexpected. Normally, no piezo-resistance is observed in *n*-type InAs[54] and other bulk semiconductors with a dominant conduction band of s-type symmetry.[55] However, similar measurements made on InAs nanowires have shown a similar piezo-resistance.[19] The effect was interpreted as a modulation of the Schottky barrier height.[39] However, the method presented here allows a precise measurement of the contact resistance as the difference between the two-probe resistance, measured across the central InAs nanowire segment, and the four-probe resistance. The changes in contact resistance are negligible in our experiment, proving that the origin of the piezo-resistance in our experiment cannot to be attributed to changes in the Schottky barrier height. We believe that the reason for the discrepancy between our experiment and current literature can be found in the sample-preparation procedure. Our method benefits from using the same fabrication processes as in traditional CMOS technology, avoids the difficulties encountered wit using less mature fabrication procedures, and allows a device performance to be achieved like that obtained in traditional CMOS substrates. From this perspective, the measurement of the InAs nanowire piezo-resistance is a good example of the potential advantages that this experimental platform can offer. However, a detailed discussion to understand the physical origin of the anomalous piezo-resistance in InAs would exceed the scope of this paper and will be given elsewhere.[56]

## Conclusion

In conclusion, we have presented a characterization platform that allows nanomaterials to be characterized as function of stress by automatically measuring the electrical transport and optical spectroscopy. The device to be tested is fabricated on the surface of a flexible substrate, which can be bent in a three-point geometry to induce uniaxial stress. Both tensile or compressive stress can be achieved and varied continuously in a range of up to 10 %, with a high resolution in axial strain up $10^{-7}$ %. We discuss the choice of materials used for the substrate and the requirements for compatibility with both the mechanics and the optical and electrical measurements. The challenges of establishing contact to the substrate even during bending are discussed, and a solution that uses off-the-shelf components is presented. The contact reliability upon bending is quantified and maximized: variations of less than 5 Ω are observed consistently during one stress cycle. The structure of the instrument control software that enables automatic measurements is presented and discussed. The software itself is made available under open-source license. To demonstrate the capability of the instrument, we characterized the piezo-resistance of an InAs nanowire device. We describe the fabrication of this device, and show the dependence of the electrical characteristics and of the Raman spectrum upon the application of stress. Uniaxial stress is shown to modulate the two-point resistance over three orders of magnitude reversibly and reproducibly, as the Raman peak shifts between 208 $cm^{-1}$ and 222 $cm^{-1}$ and is used as strain gauge. We believe that applying this methodology systematically to nanoscale devices and novel materials for electronics will help evaluate the potential of strain as a technology booster for novel nanoscale electronic devices.

## Acknowledgements

### Financial Contributions

The research leading to these results has received funding from the European Union Seventh Framework Program (FP7/2007-2013) E2Switch under Grant Agreement No. 619509 and ICT-2013-11 IIIVMOS under Grant Agreement No. 619326.

### Authors Contribution

G. Signorello and H. Riel conceived the experiment. G. Signorello designed the mechanical setup, and M. Tschudy and M. Buerge realized it. H.R. Steinauer designed and realized the electrical connectors. G. Signorello, M. Schraff and U. Drechsler designed and fabricated the substrates. G. Signorello, M. Schraff and P. Zellekens fabricated the samples. G. Signorello and H. Riel wrote the manuscript, which was discussed by and received contributions from all authors.

---

[56] G. Signorello, S. Sant, N. Bologna, M. Schraff, U. Drechsler, H. Schmid, S. Wirths, M. D. Rossell, A. Schenk, and H. Riel, submitted (2016).

## Website links

The instrument control software released under open source license can be found at the following link:

https://github.com/5igno/oedus

The references include the following links to websites: